# Capitalising the Network Externalities of New Land Supply in the Metaverse

Kanis Saengchote[1], Voraprapa Nakavachara[2,+], Yishuang Xu[3]

Version: 29 March 2023

**ABSTRACT**

When land becomes more connected, its value can change because of network externalities. This idea is intuitive and appealing to developers and policymakers, but documenting their importance is empirically challenging because it is difficult to isolate the determinants of land value in practice. We address this challenge with real estate in The Sandbox, a virtual economy built on blockchain, which provides a series of natural experiments that can be used to estimate the causal impact of land-based of network externalities. Our results show that when new land becomes available, the network value of existing land increases, but there is a trade-off as new land also competes with existing supply. Our work illustrates the benefits of using virtual worlds to conduct policy experiments.

Keywords: Metaverse, NFT, Hedonic Land Price, Network Externalities, Natural Experiment

[1] Chulalongkorn Business School, Chulalongkorn University (email: kanis@cbs.chula.ac.th).
[2] Faculty of Economics, Chulalongkorn University (email: voraprapa.n@chula.ac.th ).
[3] Manchester Urban Institute, University of Manchester (email: yishuang.xu@manchester.ac.uk ).
+ Corresponding Author.
Acknowledgement: We greatly appreciate the manual data collection work done by our research assistants, Sirima Jirathanasak and Supatsorn Kirdratanasak. The authors have no conflicts of interest to declare.

# 1. Introduction

Network externalities or network effects refer to the positive feedback effects in markets where complementary products or services form a network, and the network's value increases with its size (Katz and Shapiro, 1994). Many information-based systems, such as e-commerce and social platforms, exhibit network externalities and have received much scholastic attention recently (McIntyre and Srinivasan, 2017; Rietveld and Schilling, 2021). While this strand of research has its roots in industrial economics, this complementarity reminisces the influence of the real estate location effect, where the value of a property depends not only on its attributes but also on its surroundings. Indeed, some real estate researchers have explicitly applied the concept of network externalities to study the agglomeration effect (Capello, 2000; Burger and Meijers, 2016; Huang, Hong, and Ma, 2020). When a city becomes more connected, it grows faster; thus, urban network externalities can explain variations in regional growth rates. In this paper, our objective is to apply the concept of network externalities to land and analyse how changes in land supply can affect land value.

Land is a natural resource that is an essential productive input for many economic uses, and its value depends on a series of social, economic, legal and environmental factors across space and time (Grigsby, 1986), ranging from physical attributes and surroundings to public policies, demographic changes, and economic conditions. From the perspective of network externalities, an increase in land supply in adjoining areas can contribute to land value. While land is a finite resource that cannot be easily created, real estate developments that are urban renewal projects effectively increase land supply. However, these developments are multifaceted and occur over a long period, making it difficult to isolate the effect of network externalities with causal inference tools, particularly when the factors are predicted to have the same influence. For example, in urban renewal projects, blighted properties are often demolished, and their removal can positively impact land value (Paredes and Skidmore, 2017). Therefore, it is not possible to distinguish how much of the price increase can be attributed to network externality and neighbourhood aesthetics. We address this challenge with a virtual economy, which provides a unique environment to analyse the network externalities of land.

The advent of the Internet led to the development of virtual economies – online worlds where users can interact in various ways, including trading virtual goods for virtual currencies (Castronova, 2001). Some virtual economies allow users to possess and transfer real estate ownership. Such design became more popular as crypto assets and blockchain gained broader interest in 2020. Virtual economies had been referred to "metaverse", and the decision by Facebook to announce a name change to Meta in October 2021 further popularised this term.

In this paper, we analyse The Sandbox metaverse – a virtual gaming world built on the Ethereum blockchain where players can own and trade digital assets, including ownership of in-game land parcels, represented as non-fungible tokens (NFTs) referred to as LANDs. The Sandbox is designed to have a maximum of 166,646 (408 x 408) LAND parcels to be gradually released via public sales in waves (phases). However, the locations of the new LAND parcels and announcement dates of sales are not pre-specified, providing a series of natural experiments that can be used with an appropriate causal econometric method to isolate the effect of new land supply. Unlike urban renewal projects whose announcement dates often follow lengthy discussions and involve many changes, land sales announcements in The Sandbox cannot be easily unanticipated by participants and are not accompanied by other developments that could confound the effect. This novel research setting allows the network externalities of land to be examined more accurately.

When new land sales are announced, there are two competing forces. On the one hand, the increase in LAND supply, which can be viewed as competition, can reduce the price of existing LANDs; on the other hand, the complementary network externalities can increase the price. Using 9,920 LAND transactions in The Sandbox between December 2019 and January 2022 and the difference-in-differences hedonic price regressions with fixed effects, we find that an announcement of new LAND supply increases the prices of nearby LANDs by 8.4% in the seven days following the announcement. Our finding supports the network externalities hypothesis and shows that this effect dominates the negative price impact from increased supply.

In a subsequent analysis, we distinguish between the announcements involving multiple waves of LAND sales and those with a single wave. We find that multiple wave announcements introducing more LAND supply experience lower post-announcement treatment effect, confirming that the competition effect of LAND supply is also present. Finally, we divide the sample into periods before and after Facebook's name change announcement (which we call pre-Meta and post-Meta periods) and find that the network externalities are present in the pre-Meta period but not the post-Meta period. The inference of network externalities capitalised into LAND prices relies on the assumption that users make a rational assessment. In the face of a speculative frenzy, it is possible that users were no longer making sound decisions and were trading for short-term capital gains instead.

The rest of this paper is organised as follows. Section 2 provides the background on The Sandbox and its LAND sales. Section 3 reviews related literature and motivates the network externalities hypothesis. Section 4 describes the data sources and empirical methodology. Section 5 presents and discusses the results, and Section 6 concludes.

## 2. The Sandbox

The Sandbox is a virtual gaming world where players can own and trade digital assets such as real estate and 3D models constructed from 3D cubes (known as "voxel"). Part of the virtual environment is built on the Ethereum blockchain, a programmable blockchain that allows users to create new digital information in various forms freely but broadly referred to as "tokens". For example, an ERC-20 token is equivalent to numbers representing loyalty points, currency, or other fungible assets such as stock shares. Different token standards allow for more unique information units for developers who wish to create more heterogeneity. These are broadly referred to as non-fungible tokens (NFTs), better suited to present differentiated goods such as artwork or ownership titles.

While there have been many virtual gaming worlds (e.g., Second Life, World of Warcraft, and Minecraft), the incorporation of permissionless blockchain into the virtual world's design allows users more significant freedom over their in-game digital assets. For example, virtual real estate can be freely owned or traded by anyone who has not participated in the virtual world. This new breed of virtual worlds began gaining attention as interest in crypto assets and blockchain emerged in 2020. In its White Paper released in 2020, the developers describe The Sandbox as "a unique virtual world where players can build, own, and monetise their gaming experiences using SAND, the main utility token of the platform" with a mission of building "a system where creators will be able to craft, play, share, and trade without central control." [1]

Here, SAND is an ERC-20 token that can be used to purchase digital assets represented as NFTs, similar to in-game currency. The Sandbox often uses SAND to incentivise users' participation in the virtual world. NFTs provide unique identification numbers linked to in-game digital objects such as real estate, and developers can allow them to be transferable between users on the blockchain. In a permissionless blockchain, anyone can create an account referred to as an "address" – representing "somewhere" to send digital objects to, rather than "someone", as no identification is required to create an address. A permission blockchain is also designed so that the owner of an address can send instructions such as transfer to the information system without fear of being censored. This "permissionless" property is because the system is designed to not rely on a single administrator who decides whether to process the transaction. These unique features of permissionless blockchain allow anyone with an address on the same blockchain can freely interact with each other, which is what the phrase "trade without central control" in The Sandbox's mission means.

[1] Source: https://installers.sandbox.game/The_Sandbox_Whitepaper_2020.pdf

While The Sandbox allows users to generate voxel content represented as NFTs called "ASSETS freely", in-game real estate is created by The Sandbox. It is also defined as NFTs and called "LAND". According to the White Paper, LANDs represent "physical parcels of The Sandbox Metaverse". LAND ownership allows users to freely host content, create and monetise self-designed games, or arrange other commercial activities such as storefronts or event space, making it similar to traditional real estate. The map of The Sandbox is a 408 x 408 grid, and each LAND NFT is identified by coordinates, so the intended maximum size of the map is 166,464 parcels. Figure 1 illustrates a snapshot of The Sandbox's map as of 24 August 2022. The coloured regions are LANDs that The Sandbox has already created. Users can buy separate parcels of LAND (lot size = 1) or purchase adjacent parcels together as bundles (lot size > 1) or ESTATEs (S: 3x3, M: 6x6, L: 12x12, XL: 24x24).

**Figure 1: Map of The Sandbox**

This figure is the screenshot of the Sandbox map captured on 24 August 2022. The intended size of the map is 166,464 squared parcels formed by 408 x 408 grids, and each grid is represented by a non-fungible token (NFT) called LAND. The coloured regions are LANDs already created by The Sandbox, while the black regions are the areas where LANDs have not been released.

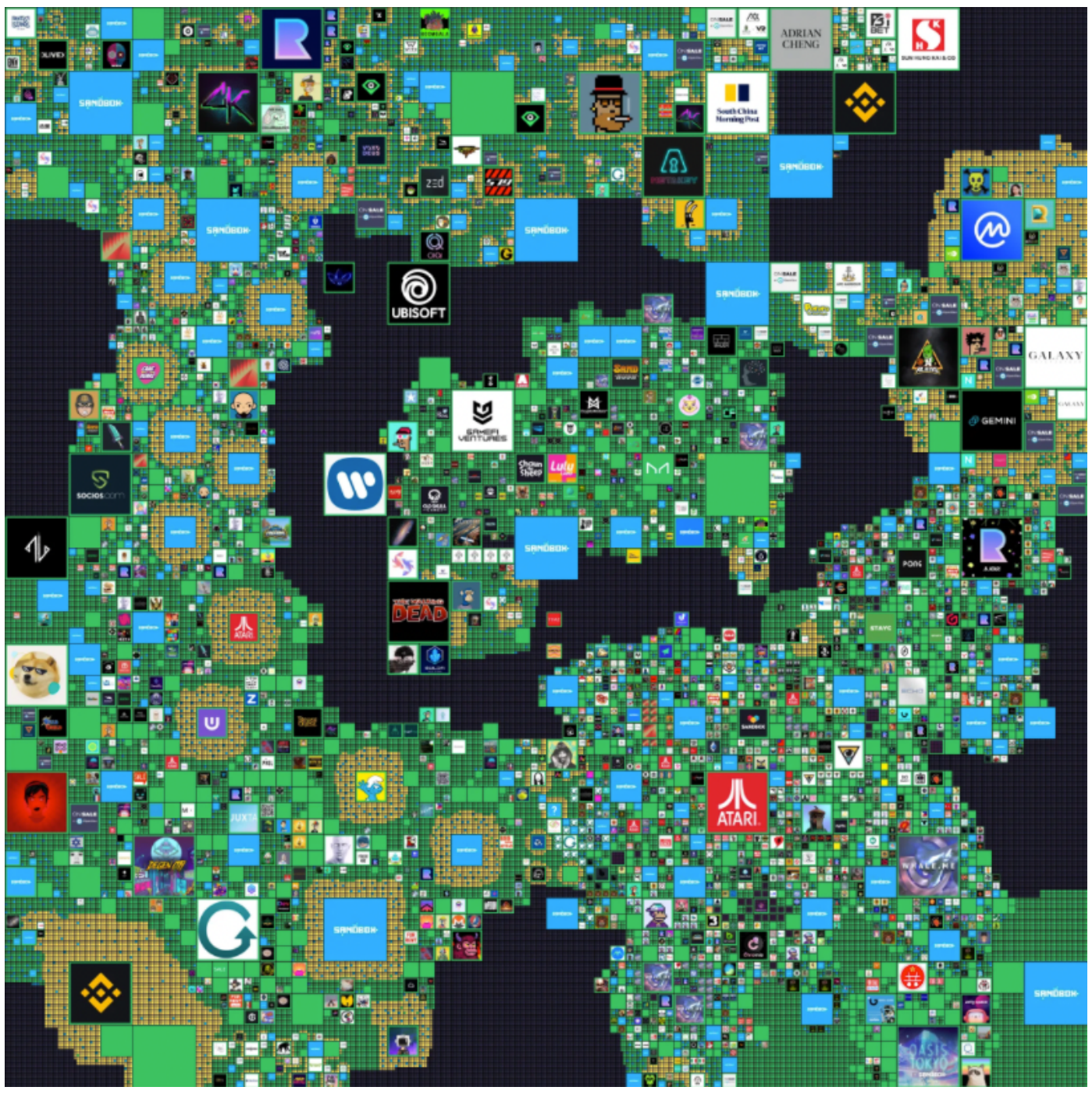

The black regions are LANDs that have not yet been created, which The Sandbox makes available in "waves". Before the actual sale date of each round, the developers would announce details such as locations of the parcels to be offered, prices for different types, and when users can buy them. We manually collected the details of LAND sales between the launch and December 2021, during which there were 34 waves. For some waves, The Sandbox may partner with projects or celebrities, such as Snoop Dogg in waves 32 and 33, who can act as "anchor tenants" for the wave. For example, the partner for the Public Sale Wave 1 announced on 26 January 2021 is CoinMarketCap, a crypto asset data aggregator. The wave contains 1,740 LAND parcels, 1,200 of which are offered as Premium LAND that bundle virtual character NFTs and LAND and sold for 4,683 SAND each. Users could buy LAND on 11 February, approximately two weeks after the announcement. For each wave, what constitutes Premium LAND can be different: for some waves, it could be an NFT bundle; for others, it could be location-based. Regular LAND is always offered at 1,011 SAND, and Premium LAND at 4,683 SAND.

These sales are considered primary market sales, and developers of The Sandbox are the beneficiaries of such sales. LAND owners may later choose to resell their holdings in secondary markets at any time. While permissionless blockchain allows users to transact directly and freely, users often use third-party aggregators such as OpenSea or LooksRare, a peer-to-peer marketplace for NFTs that can help with the search friction and price discovery process. Details of the waves are reported in Table 1, and the geographical distribution of LAND sale waves is illustrated in Figure 2.

### Table 1: The Sandbox's LAND Sale Waves

This table provides information on The Sandbox's LAND sales until December 2021. LANDs are created (minted) and sold in waves (rounds). Before each wave, The Sandbox creators announce the details of the LANDs that will be sold for that wave, including the locations and the sale date. Sale dates shortly occurred before announcement dates, and multiple waves can be jointly announced simultaneously. The details of LAND sales are manually collected from The Sandbox's website.

| Group | Wave | Wave Name | Ann. Date | Sale Date | LAND offered |
|---|---|---|---|---|---|
| 1 | 1 | Presale Round 1 | 22-Nov-19 | 5-Dec-19 | 3,096 |
| 2 | 2 | Presale Round 2 | 10-Jan-20 | 2-Feb-20 | 6,192 |
| 3 | 3 | Presale Round 3 | 12-Mar-20 | 31-Mar-20 | 12,384 |
| 4 | 4 | MoonSale#1 | 26-May-20 | 4-Jun-20 | 1,458 |
| 4 | 5 | MoonSale#2 | 26-May-20 | 11-Jun-20 | 1,458 |
| 4 | 6 | MoonSale#3 | 26-May-20 | 16-Jun-20 | 711 |
| 4 | 7 | MoonSale#4 | 26-May-20 | 23-Jun-20 | 1,458 |
| 4 | 8 | MoonSale#5 | 26-May-20 | 30-Jun-20 | 1,458 |
| 4 | 9 | MoonSale#6 | 26-May-20 | 2-Jul-20 | 711 |
| 5 | 10 | Presale 4 Round 1 | 22-Jul-20 | 4-Aug-20 | 4,536 |
| 6 | 11 | Presale 4 Round 2 Wave 1 | 1-Sep-20 | 15-Sep-20 | 1,171 |
| 6 | 12 | Presale 4 Round 2 Wave 2 | 1-Sep-20 | 16-Sep-20 | 1,095 |
| 6 | 13 | Presale 4 Round 2 Wave 3 | 1-Sep-20 | 17-Sep-20 | 1,046 |
| 6 | 14 | Presale 4 Round 2 Wave 4 | 1-Sep-20 | 18-Sep-20 | 1,224 |
| 7 | 15 | Presale 4 Round 3 | 29-Oct-20 | 12-Nov-20 | 19,200 |
| 8 | 16 | Public Sale Wave 1 | 26-Jan-21 | 11-Feb-21 | 1,740 |
| 9 | 17 | Public Sale Wave 2 | 19-Feb-21 | 25-Feb-21 | 2,302 |
| 10 | 18 | Public Sale Wave 3 | 9-Apr-21 | 14-Apr-21 | 3,323 |
| 11 | 19 | Public Sale Wave 4 | 22-May-21 | 27-May-21 | 1,777 |
| 12 | 20 | Summer Jam Wave 1 | 19-Jun-21 | 24-Jun-21 | 1,602 |
| 12 | 21 | Summer Jam Wave 2 | 19-Jun-21 | 1-Jul-21 | 2,111 |
| 12 | 22 | Summer Jam Wave 3 | 19-Jun-21 | 8-Jul-21 | 472 |
| 12 | 23 | Summer Jam Wave 4 | 19-Jun-21 | 15-Jul-21 | 1,381 |
| 12 | 24 | Summer Jam Wave 5 | 19-Jun-21 | 22-Jul-21 | 1,614 |
| 13 | 25 | Summer Jam Wave 6 | 30-Jul-21 | 5-Aug-21 | 1,019 |
| 13 | 26 | Summer Jam Wave 7 | 30-Jul-21 | 12-Aug-21 | 1,058 |
| 14 | 27 | The Walking Dead Wave 1 | 21-Aug-21 | 26-Aug-21 | 920 |
| 14 | 28 | The Walking Dead Wave 2 | 21-Aug-21 | 2-Sep-21 | 597 |
| 14 | 29 | The Walking Dead Wave 3 | 21-Aug-21 | 9-Sep-21 | 513 |
| 14 | 30 | The Walking Dead Wave 4 | 21-Aug-21 | 16-Sep-21 | 802 |
| 15 | 31 | Pororo the Little Penguin | 9-Sep-21 | 23-Sep-21 | 892 |
| 16 | 32 | Snoop Dogg Wave 1 | 23-Sep-21 | 2-Dec-21 | 192 |
| 16 | 33 | Snoop Dogg Wave 2 | 23-Sep-21 | 16-Dec-21 | 203 |
| 17 | 34 | The Quan | 3-Nov-21 | 4-Nov-21 | 254 |

**Figure 2: Map of The Sandbox's LAND Sale Waves**

This figure displays the geographical distribution of The Sandbox's LAND sales by wave, each shaded with a different colour for easy identification. There is no clear spatial pattern on where the next wave will be. The coloured regions are LANDs that have already been created by The Sandbox, while the black regions are the areas where LANDs have not been released.

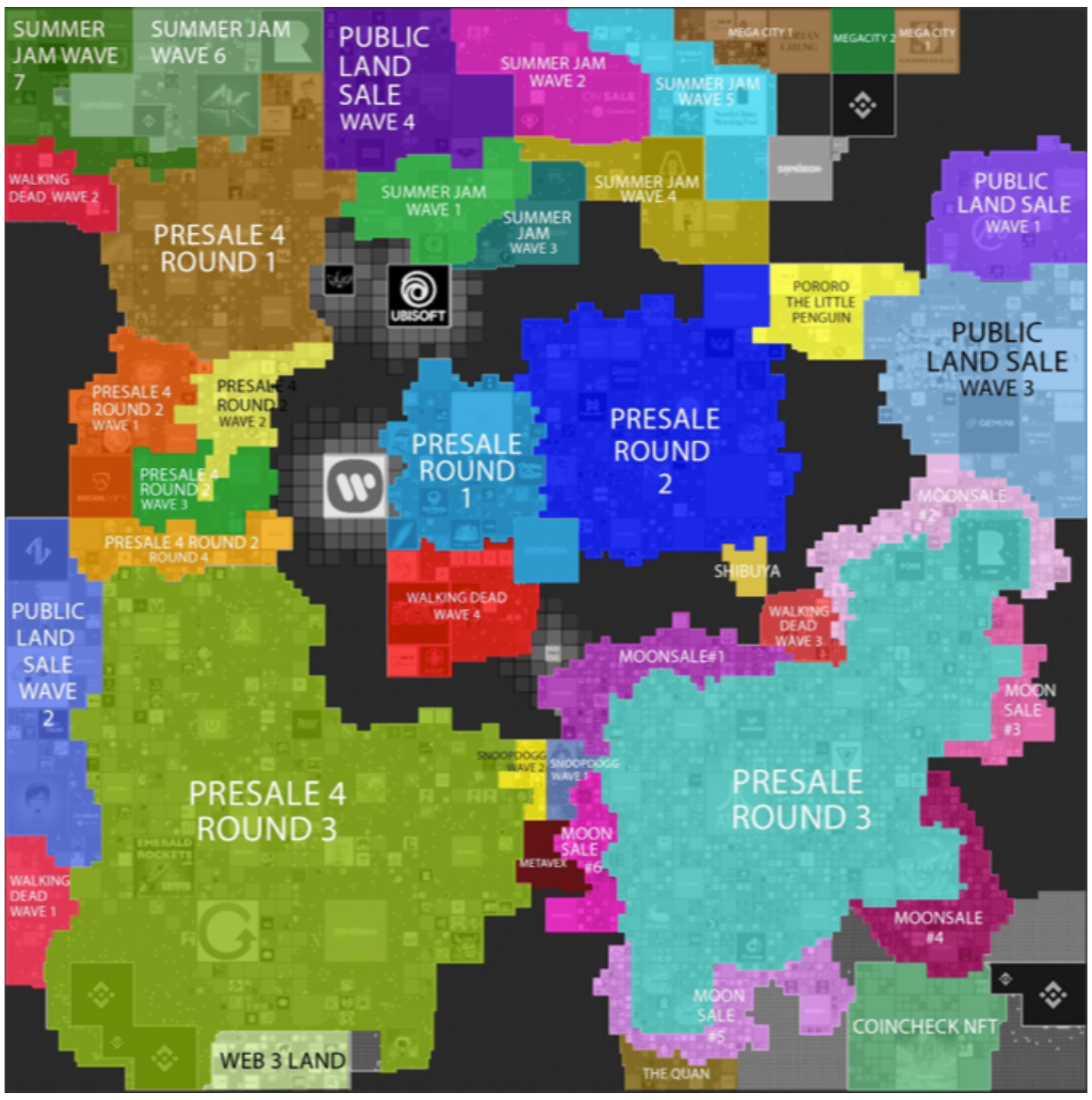

To the best of our knowledge, users do not know in advance when the next wave will occur, and there is no noticeable pattern among the 34 waves of where the next region offered will be. Consequently, these unanticipated events provide a series of natural experiments that can be used to analyse how users respond to exogenous changes in land supply, which is challenging to do in the real world.

## 3. Literature Review and Hypothesis Development

This section reviews the literature related to land value and issues that metaverse real estate can offer a unique research setting. Then, we develop the testable hypotheses before describing the data sources and empirical methodology in the next section.

### 3.1 Determinants of Land Value

Land is often viewed as a heterogeneous good based on its characteristics and uses. Because land is a natural resource that is a productive input for many economic benefits, its value depends on many factors. For example, Grigsby (1986) suggests that land prices are affected by social, economic, legal and environmental factors over time and across space. Due to the social nature of The Sandbox and the potential to conduct commercial activities in the metaverse, we focus on research related to urban land. To formalise the framework, we follow Potepan (1996), who classifies factors that affect urban land prices as endogenous and exogenous. The endogenous factors are about the physical attributes of land, such as its characteristics, size, location, and surroundings, while the exogenous factors are about various forces that influence when and how land can be used. The influences can come in many forms, such as public policies, demographic changes, economic conditions, and investment decisions.

For urban land, proximity to employment opportunities and amenities and ease of commute via different modes of transportation are essential drivers of value (e.g., Mulley, 2013). While larger parcels of land have higher prices, their price per unit of land can be lower. Lin and Evans (2000) argue that bigger plots receive lower prices because they have fewer structural improvements per unit area and are difficult to access.

Amongst the exogenous factors, regulatory restrictions enacted as parts of public policies have a vital role but can vary greatly depending on the locales. Consequently, a fundamental challenge with such studies is that the impact can depend on the case's circumstances. For example, Ihlanfeldt (2007) concludes that heavy land regulation has a negative effect on local land prices based on the case of cities in Florida, while Pollakowski and Wachter (1990) analyse land use regulations in Maryland and find that constraints may not lead to the rising land price as the market environment limits their impact. Moreover, the interconnection between real estate and other economic activities often means other public policies, such as environmental regulations (Chattopadhyay, 1999) and industrial policies (Fehribach et al., 1993), can affect land value. The inability to isolate a specific factor's impact from others makes statistical inference of empirical real estate research challenging.

Demographic factors such as population growth and movements have also received significant attention for affecting the change on the demand side of land. For example, Rose (1989) finds that the land price is positively associated with population or its growth rate, while Rye (2011) and Saiz (2007) indicate that immigrants also have an impact on land price as they contribute to the total population. For economic factors, regional income and real interest rates are factors that can influence demand for real estate (Potepan, 1996; Wang, 2009), with linkage through the collateral channel (Bernanke, Gertler, and Gilchrist, 1999; Almeida, Campello, and Liu, 2006; Liu, Wang, and Zha, 2013).

Compared to other exogenous factors, the investment factors have not received as much attention until the 1990s, as noted in a survey study by Sirmans and Worzala (2003). Investment factors can take place in many forms. For example, foreign demand for a country's real estate could increase property prices. Badarinza and Ramadorai (2018) show that "flights to safety" can lead to an increase in real estate demand in global cities such as London and New York, while Rodríguez and Bustillo (2010) analyse the case of Spain, which is more driven by demand for tourism. Both studies find that increased external demand has a lasting impact on property prices. The international capital flows can also affect the financial market conditions of a country and influence real estate investment decisions, as documented for the United States by Favilukis et al. (2013).

### 3.2 Hedonic Pricing Model

Economists often assume that price reflects the willingness to pay. Suppose one surmise that objectively measured characteristics can completely describe a differentiated good. Under this view and some assumptions about how people act and the market behaves, observed market prices of a differentiated good can be used to infer the prices of various characteristics. This idea is often attributed to Rosen (1974), who calls these "hedonic" prices. The use of hedonic prices exists in many contexts, including real estate. For example, the S&P Case-Shiller house price index based on Case and Shiller (2003) is also based on this idea. As a property includes land and structure components, land value capitalises the market value of a location in the absence of the structure (Davis and Heathcote, 2007; Haughwout, Orr, and Bedoll, 2008), and the flexibility that land ownership provides can give rise to real option value (Yamazaki, 2001). Examples of applications in land pricing include Clapp (1990), Bostic, Longhofer, and Redfearn (2007), Spinney, Kanaroglou, and Scott (2011), and Costello (2014).

Hedonic pricing equations often take the form of Equation 1, where $y_{it}$ is the log dollar price of property $i$ at time $t$ and $X_{it}$ is the potentially time-varying characteristics of the property. The vector $\gamma$ contains the hedonic price estimates per unit of objectively measured characteristics. The hedonic pricing model is the workhorse of empirical real estate research and can be used in various ways. Often, researchers may include time fixed effects $\tau_t$ to address exogenous factors that affect the overall market, such as macroeconomic conditions, which result in a general rise in property prices during that period. The estimated fixed effects would represent average prices that cannot be explained by the hedonic pricing model and can be used to construct real estate price indices.

$$y_{it} = \alpha + \gamma' X_{it} + \tau_t + \varepsilon_{it} \tag{1}$$

### 3.3 Economics of New Land Supply

Land is considered a finite resource that cannot be easily created. However, not all land is usable, so in the context of our paper, an increase in land supply refers to developable land that participants could use. Like other goods and services, the equilibrium land price can be viewed in the context of demand and supply. When demand remains unchanged, the increase in the supply of developable land would decrease land price and vice versa.

Indeed, researchers have long looked at the effect of restrictive land supply on house prices, such as in South Korea (Hannah et al., 1993) and Hong Kong (Peng and Wheaton, 1994). Urban policies restricting land supply, such as an urban growth boundary (UGB), can increase land prices, but the impacts vary. For example, Mathur (2014) finds that UGB in King County, Washington, positively impacted land prices, while Buxton and Taylor (2011) find no impact for UBGs in Melbourne.

It is more common for land supply to be restricted than increased. However, policies such as urban renewal projects can be considered an increase in developable land, and they often positively impact the value of existing properties. For example, Rossi-Hansberg, Sarte and Owens III (2010) develop a model of housing price that capitalises benefits into land value and find that residential urban revitalisation programs implemented in Virginia had a positive impact on land prices, while Chau and Wong (2014) find similar positive effects in Hong Kong. These positive effects can arise from many factors. For example, Paredes and Skidmore (2017) argue that when blighted properties in Detroit are demolished, the adverse effects of dilapidation on the neighbourhood are removed. Thus the net benefit could be positive. The same argument is made by Chau and Wong (2014) for urban renewals in Hong Kong.

The positive spill over effect can arise because a portion of land value is also derived from its surroundings, benefiting from "network effects". A review article by Katz and Shapiro (1994) discusses examples of complementary products and services with little or no value in isolation but generate value when combined with others. Markets with this positive feedback effect are referred to as systems markets. Such systems benefit from "network effects" or "network externalities", where the network's value increases with the network's size. In the context of urban studies, the network externalities come from agglomeration economies. The wider concept of agglomeration economies is explained by Alonso (1973) as the effect of 'borrowed size', which can be interpreted as the fact that cities can 'borrow' the functions of other cities because of the physical proximity or other relationships (Boix and Trullén, 2007;

Camagni and Salone, 1993). The urban network externalities lead to a particular type of agglomeration, which indicates that the economy of a city can benefit from being embedded in networks and/or connected well to other cities.

The increase in the supply of developable land surrounding a plot of land can increase the network effect of that land. This complementarity reminisces the agglomeration effect of urbanisation (Capello, 2000; Burger and Meijers, 2016; Huang, Hong, and Ma, 2020) and provides another source of explanation for why increases in land supply can complement the value of neighbouring land. However, it is unclear whether the impact observed in studies of urban revitalisation programs is attributed to land-based network externalities or other amenities.

### 3.4 Virtual Real Estate

Despite virtual economies existing for many years, there is limited virtual real estate research in the same vein as traditional real estate research. Castronova (2001) is among the first to use the phrase "virtual economy" to refer to economies in online games, where artificially scarce goods and currencies can be traded for real money. In the context of the digital world, where virtual goods are information that can be freely created, any scarcity must be artificially induced. Under this similar definition, a 2011 World Bank working paper by Lehdonvirta and Ernkvist (2011) points out several commercial opportunities in the virtual economy, such as improving the visibility of an online store and the ability for users to create virtual goods. However, there is no explicit reference to virtual real estate, as not all virtual economies allow users to have real estate ownership.[2] Among research examining virtual real estate, Xiao-lin et al. (2010) collect and analyse data from SecondLife.com and find that virtual land prices correlate with real-life locations' value.

Data availability presents another obstacle in conducting virtual real estate research, as virtual real estate is essentially private, digital information created by developers. Thus their transactions are not required to be publicly recorded like traditional real estate. With the advent of permissionless blockchain and the representation of virtual real estate ownership as NFTs, researchers gain better visibility into the transactions of virtual economies. For example, Dowling (2022) analyses LAND NFTs of Decentraland, another blockchain-based virtual world similar to The Sandbox, and finds that the time series averages of LAND prices exhibit evidence of market inefficiency. Goldberg, Kugler and Schär (2021) also investigate Decentraland to analyse whether locational preferences affect real estate prices as this virtual world has negligible transportation costs. They find that land close to popular landmarks and parcels with memorable coordinates receives higher bid prices. For The Sandbox, Nakavachara and Saengchote (2022) find that users who buy LAND using SAND that appreciated rapidly pay at higher dollar prices. They attribute the findings to a different unit of account that buyers use to assess the price they end up paying. Thus LAND appears cheaper to them when paid in SAND than other cryptocurrencies.

[2] Second Life – one of the earliest virtual economies – allows users to transfer ownership of their virtual real estate. See, for example, https://marketplace.secondlife.com/products/search?search%5Bcategory_id%5D=852.

While research on real estate in virtual economies remains limited, users in these virtual worlds exhibit behaviour that can be explained by economic theory. Thus, we will use the natural experiments from The Sandbox virtual economy to gain insights into the complex interaction of increased land supply and the real estate market.

### 3.5 Hypothesis Development

The challenge of identifying land-based network externalities lies in the research setting that is not conducive to using causal inference tools. Researchers often use event studies to measure causal, price-based impacts of real estate attributes. As discussed earlier, increases in new land supply occur in the context of new real estate developments. However, developments typically occur after lengthy planning processes involving stakeholder consultations and formal approvals and are subject to uncertainty. Consequently, announcements of new developments do not provide an exogenous shock required for event studies. In this case, the price impact needs to be measured over a long horizon, so price increases detected in statistical analyses could be due to other unobserved changes in the neighbourhood that coincide with new land developments. This problem is often referred to as the omitted variable bias. It is also possible that the new developments occur because the area has been trending, so the direction of causality can be reversed.

Another challenge is isolating the impact of land-based network externalities from the multiple forces that affect land value. New developments involve several changes that can have the same directional impact on prices, but researchers only observe one price that incorporates all forces, positive and negative. An inference can be made about the net effect in a research setting with competing forces. However, with reinforcing forces, the inability to explicitly measure the magnitude of the force poses an identification challenge, which can be viewed as a limitation of the hedonic pricing model.

These examples are endogeneity issues that challenge the causality of statistical results. A perfect natural experiment would involve an exogenous increase in land supply that is not accompanied by any other changes and is unrelated to past price changes, which is rare - if not impossible - to find in traditional real estate. The virtual real estate market of The Sandbox provides a unique opportunity to pursue this topic, given its similarity to traditional real estate and the ability of developers to exogenously increase land supply.

The new LAND sales/supply at the Sandbox have been unpredictable. In addition, unlike urban renewal projects, the LAND sale does not change the attributes of existing LANDs. Neither are there environmental issues such as pollutants or crime in virtual real estate, so there are fewer potential sources of variations at The Sandbox. Therefore, this research setting can be considered a natural experiment, allowing us to analyse land-based network externalities.

Similar to the traditional real estate, metaverse land prices can be influenced by endogenous factors such as location, size, and proximity and exogenous factors such as economic, demographic, and investment. Virtual amenities can be new landmarks, themed developments, or anchor residents, such as celebrities collaborating with the developer in each sale wave, as described in Section 2. Other factors, such as regulation and land use, are less vital for The Sandbox, built on a permissionless blockchain designed to encourage freedom of interaction. In addition, the Sandbox does not enforce any zoning restrictions. For proximity to transportation, The Sandbox allows users to teleport freely to locations, making logistical considerations less critical in this metaverse.

If some endogenous factors can be assumed to be invariant over time, and exogenous factors assumed to be common across all the virtual economy at a given point in time, the difference-in-differences (DiD) method with fixed effects can be used to control for all these factors, and value of amenities that newly created LAND provides can be isolated. Furthermore, The Sandbox virtual economy can be considered a network that competes against other networks, such as Decentraland, so the inclusion of time fixed effects can account for the general increase in network value arising from increased membership in the virtual economy.

The natural experiment is an announcement of a new LAND sale, and we analyse the impact of the new supply on existing LAND. From the discussion in Section 3.3, two competing forces could affect the prices of existing LAND: on the one hand, an increased LAND supply can be viewed as competition and thus suppress prices; on the other hand, the complementary network externalities can increase the value. The result from the DiD method would represent the localised network externalities capitalised into transaction prices. The closer to the new LAND, the greater the externalities. We will discuss the DiD method further in Section 4.2.

# 4. Data and Methodology

## 4.1 Data

Data on permissionless blockchains like Ethereum can be transparently inspected, but not all relevant information is recorded on the blockchain. Because space is scarce, developers often record just enough information on-chain and supplement it with additional data stored elsewhere. For the case of The Sandbox, the LAND NFTs that represent virtual real estate ownership only contain unique identification numbers, which must then be linked to attributes such as location data maintained by The Sandbox. We retrieve all instances of LAND transfers between addresses and the corresponding payments made on Ethereum and process them to generate data real estate transactions. [3] We exclude all transfers that do not include payments, which may be gifted transfers or involve off-chain payments agreed privately between counterparties. We assume that on-chain payments reflect the total payment amount. This assumption is reasonable because more than 99% of the transfers are arranged via smart contracts of OpenSea, the leading marketplace for NFT listings.

Our sample starts from 5 December 2019 to 28 January 2022, before The Sandbox migrated its LAND contract to a new address. LAND transfers involving on-chain payments include primary sales by the Sandbox and secondary sales, most of which are conducted via OpenSea. For our research question, only secondary sales are analysed. The location and map data are retrieved from The Sandbox's data API and matched to the identification number (ID) of each LAND NFT. Lastly, the Sandbox's LAND sale announcements and the corresponding NFT IDs and locations for each wave of sale until December 2021 are manually collected from The Sandbox's official announcement websites. [4] The dates of LAND sale waves are reported in Table 1. We group the waves by common announcement dates, resulting in 17 groups from the 34 waves.

---

[3] The Ethereum address for the LAND contract is '0x50f5474724e0ee42d9a4e711ccfb275809fd6d4a'.

[4] They are posted on the Sandbox's official medium pages: https://medium.com/sandbox-game

Blockchain data records the ID of LAND NFTs and the amount of cryptocurrencies transferred between addresses. While permissionless blockchain means users can freely transact, most rely on OpenSea, which restricts the choices to popular and liquid cryptocurrencies. More than 99% of the transactions are settled in ETH, SAND, and wETH, so we restrict our analysis to these three cryptocurrencies. In addition, Nakavachara and Saengchote (2022) find that users appear to make decisions based on the dollar value, so we supplement the on-chain data with exchange rates retrieved from CoinGeckco's data API. The dollar prices are winsorised at thresholds of 0.1% and 99.9% to limit the influence of outliers.

### 4.2 Methodology

In financial research, the event study method pioneered by Fama et al. (1969) is used to analyse the impact of unanticipated news on asset prices.[5] The idea is that an asset's fundamental value can change when new information relevant to the asset is revealed. Therefore, rational investors would incorporate this new information and revise their willingness to pay. This release of information can come in many forms (e.g. announcements) and is generally referred to as an event. An event study is conducted on time series price data observed at a relatively high frequency by comparing the observed prices after the event to counterfactual prices during the same period estimated using prices before the event. The difference between the observed and predicted prices can be interpreted as a causal impact if the event is exogenous. The fewer changes occurring at the event, the easier to attribute the result.

For homogeneous and frequently traded assets such as stocks, time series price data can be easily obtained. However, real estate assets are heterogeneous and illiquid. Hence, researchers rely on a related technique that utilises transaction data of multiple assets across different points in time. Rather than using the pre-event prices to estimate the counterfactual prices of the same asset, a group of assets unaffected by the event are used as the control group and compared against the treatment group. By comparing the treatment and control groups across different points in time, we can control for common factors that affect both groups and isolate the treatment effect. The differences across time and treatment are why this methodology is known as the difference-in-differences (DiD) method, which allows a causal inference to be made in a research setting with sporadic measurements of many units of analysis.

[5] See Armitage (1995) for a survey on the variants of event study methods and their performance.

The treatment effect should depend on the LAND's proximity to the new supply, so distance is the first difference. The second difference is pre- and post-announcement. This method is used in traditional real estate in conjunction with the Rosen (1974) hedonic pricing model to uncover the value of an amenity such as pollution (Bajari, Fruehwirth, and Timmins, 2012; Currie et al., 2013) and supermarkets (Pope and Pope, 2015; Saengchote, 2020). As discussed in Section 2 and Section 3.5, the announcements of LAND sales are challenging to anticipate both temporally and spatially, so this exogeneity allays potential concerns over the validity of the empirical design.

Figure 3 illustrates how the first difference works using Public Sale Wave 1 (group 8, wave 16 on Table 1 and the first wave in our sample), announced on 26 January 2021, as an example event. The red region contains the new LANDs that will be created and available for sale on 11 February 2021. The blue regions are LANDs that have already been created from previous waves and are thus available for development prior to 26 January 2021. The white regions are not yet available as of 26 January 2021. LANDs in the blue regions that have been minted can be freely traded on the secondary markets at any time.

**Figure 3: Illustration of the Difference-in-Differences Method (Treatment vs Control)**

This figure illustrates the treatment (first difference) in the difference-in-differences method using Public Sale Wave 1 as the event of interest. The announcement date was 26 January 2021, and the sale date was 11 February 2021. The red area contains the new LANDs that will be created for this round. The blue areas are where LANDs have already been created before this announcement. The white areas are the areas that are not yet available. LANDs in the blue areas can be traded in the secondary markets, and these are the transactions of interest. For the discrete treatment, our treatment group comprises LAND transactions near the red area within the grey line (thus, near), while our control group comprises transactions outside the grey line. For continuous treatment, we use the log of the calculated Euclidean distance.

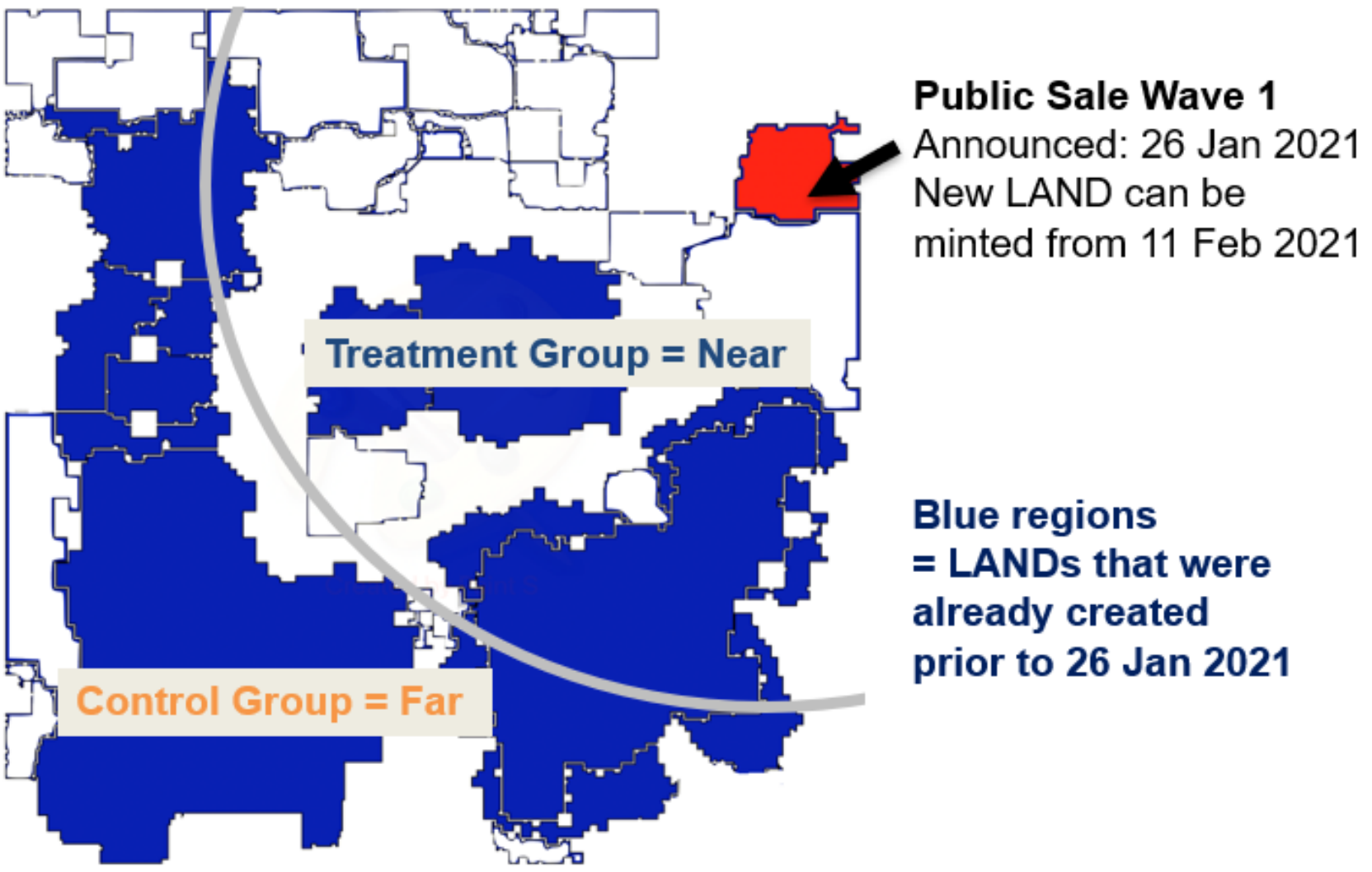

We can classify the treatment groups in two ways: the first is discrete, where LANDs closer to new parcels will be classified as the treatment group, while the second is continuous, which uses actual distances. As standard in real estate research, we define distance as the Euclidean distance (the length of a straight-line segment) between two points. The first point is the coordinates of the transacted LAND, and the second is the nearest coordinates of the new, red-shaded region. The red region contains many parcels, so the shortest distance corresponds to the closest LAND on the boundary of the new region. Under this definition, the same LAND will have a different distance for each wave, as it is measured relative to the newest parcels to be created.

LANDs can be sold in single parcels (lot size = 1) or bundles (lot size > 1). The Sandbox allows multiple parcels to be jointly transacted, and these bundled parcels are referred to as ESTATEs. For bundled transactions, we compute the minimum distance across all bundle parcels to the new region. However, a single transaction (where only a single price is observed) may involve LANDs from various locations across the map. We exclude such transactions because the relevant distance cannot be calculated.

The discrete classification threshold is illustrated by the grey line in Figure 3, which separates LANDs into near (treatment group) and far (control group). In contrast, the continuous classification uses the log of the calculated Euclidean distance to allow for more granularity. Based on the announcement dates presented in Table 1, there is, on average, one announcement per month but with no regularity. We restrict our analysis to transactions within seven days of each announcement to ensure that consecutive announcements' pre- and post-event windows do not overlap. In the context of Public Sale Wave 1, the transactions included in the sample are between 19 January and 2 February, seven days before and after the announcement date of 26 January, as illustrated in Figure 4.

**Figure 4: Illustration of the Difference-in-Differences Method (Pre vs. Post)**

This figure illustrates the pre- and post-event periods (second difference) in the difference-in-differences method using Public Sale Wave 1 as the event of interest. The announcement date was 26 January 2021, and the sale date was 11 February 2021. The pre-event period spans seven days before the announcement date, while the post-event period spans seven days after the announcement date. Based on the dates collected in Table 1, this definition will not result in any overlap between announcements.

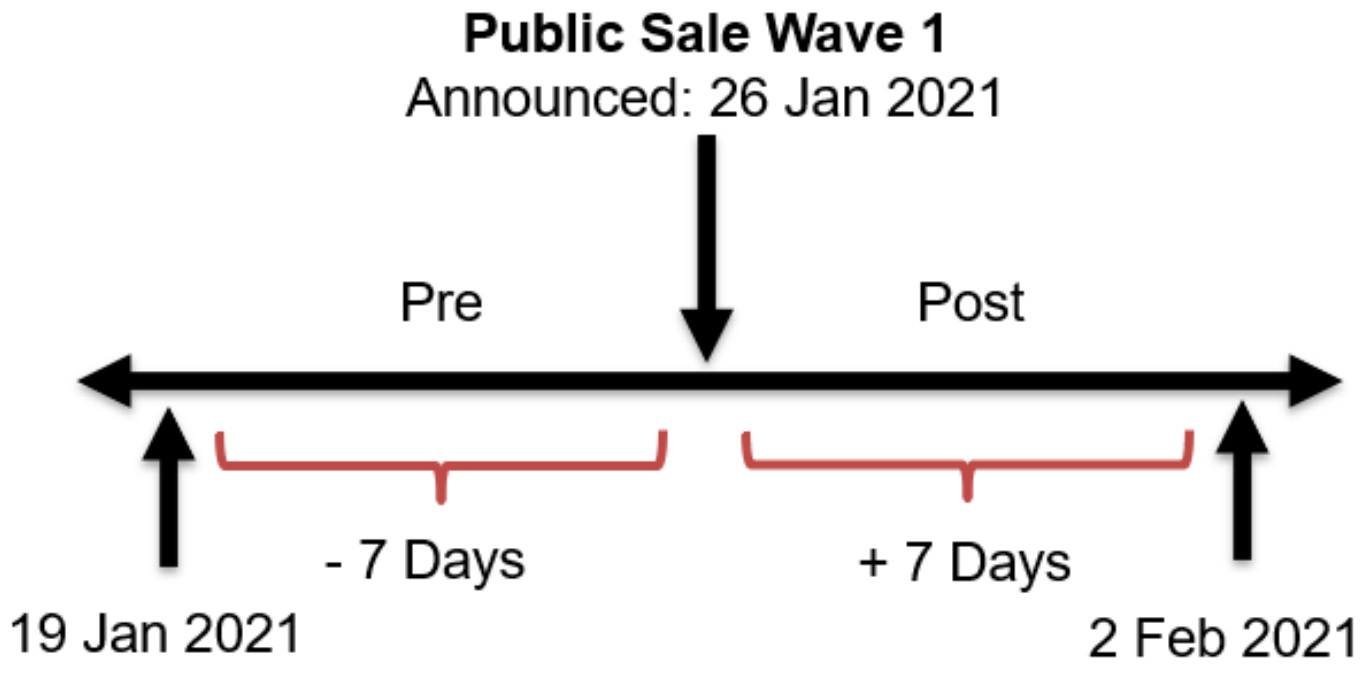

The baseline model of the DiD method is shown in Equation 2, where $y_{it}$ is the log dollar price of LAND $i$ sold at time $t$. The $Near_i$ dummy variable indicates whether LAND is near the newly announced but not-yet-available region. The $Post_t$ dummy variable indicates whether LAND is sold after the announcement dates. Finally, $X_{it}$ is a vector of control variables, including the log of lot size (number of parcels), a dummy variable of whether the Sandbox marked the LAND as Premium LAND upon sale, and the log of daily Bitcoin price in US dollars. LANDs that were first offered by The Sandbox as premium tend to be priced higher, so we include the variable to control and Bitcoin price proxies for general crypto assets conditions, which may affect the willingness to pay of users.

$$y_{it} = \beta_1 Post_t + \beta_2\ Near_i + \beta_3 Post_t \times Near_i + \gamma' X_{it} + \alpha_w + \tau_t + \varepsilon_{it} \qquad (2)$$

Nakavachara and Saengchote (2022) study the willingness to pay among The Sandbox's users and find that the choice of cryptocurrencies used to settle payments can affect the dollar price. To control for this effect, we also include dummy variables indicating whether the transactions are settled in SAND (the Sandbox's cryptocurrency) or in wETH (a tokenised version of Ether) in an extended model, with ETH (the Ethereum blockchain's native coin used to pay transaction costs) as the omitted category.

In the baseline specification, we include the time fixed effects, which we define weekly. In more restrictive specifications, we define the fixed effects daily, accounting for more variations than the inclusion of daily Bitcoin prices. In subsequent specifications, we also include the wave fixed effects αw to control for location-based heterogeneity of LANDs within a wave. Because our study aims to analyse proximity-related externalities, the inclusions of these cross-sectional and time series fixed effects provide tight controls for potential omitted variable bias. The DiD treatment effect would be identified by variations of prices based on distances within a locale, as the fixed effects would control for the general differences across locales and macro-level variations that change over time.

Under our land-based network externalities hypothesis, announcements of LAND sales would increase the prices of existing LANDs in the neighbourhood. Therefore, from Equation 2, we expect $\beta_3$ to be positive.

### 4.3 Summary Statistics

In this section, we discuss the summary statistics of The Sandbox data – particularly the distances – so the threshold for $Near$ can be motivated. We begin with the 40,958 unique transactions retrieved from the Ethereum blockchain from 5 December 2019 to 28 January 2022 and narrow it down to seven days before and after announcements, resulting in 10,974 transactions. Because the DiD method essentially computes the conditional averages across the four groups (near and far, pre and post), it is essential to have sufficient observations in each category for the estimation to have statistical power. Therefore, while we have gathered data for 34 waves, we begin our analysis from Public Sale Wave 1 (No. 16 in Table 1) to Snoop Dogg Wave 2 (No. 34 in Table 1). This screening leaves 9,920 unique transactions in our sample.

The summary statistics of the transactions are grouped by common announcement dates and reported in Table 2. The average price of LAND across all groups is $3,341.85, but group-level averages show that average prices increased until the end of 2021. The average lot size is

1.43 parcels, and about 6.2% of the LANDs are premium LANDs. More than 80% of transactions were settled in ETH, followed by 13.9% in wETH and 5.4% in SAND. The average distance to the new region is 194.96 units, with a median of 194.46. However, as illustrated by the histograms in Figure 5, the distribution varies by announcement group. As reported in Table 2, the median distance calculated by group can be as low as 67.2 and as high as 293.28. At the time of each announcement, the geographical shape of The Sandbox is different, so we define $Near$ based on the median for each group. When multiple new developable regions are released in a single announcement, we use the nearest distance between the LAND and all possible new regions.

**Table 2: Summary Statistics**

This table provides the summary statistics of the data. We begin with the 40,958 unique transactions retrieved from the Ethereum blockchain from 5 December 2019 to 28 January 2022 and narrow it down to seven days before and after announcements, resulting in 10,974 transactions. Our analysis begins from Public Sale Wave 1 (No. 16 on Table 1) to ensure enough LAND transactions in the pre-announcement period. The final sample comprises 9,920 unique transactions across ten groups, where a group represents a single announcement which may involve multiple waves. The discrete treatment variable is defined based on the median distance in each group.

| | | Price | | Distance | | | | | |
|---|---|---|---|---|---|---|---|---|---|
| Group | Obs | Mean | Std Dev | Mean | Median | Lot size | Premium | SAND | wETH |
| 8 | 252 | 856.51 | 1,402.17 | 289.97 | 293.28 | 4.12 | 0.0% | 3.6% | 6.3% |
| 9 | 463 | 1,492.21 | 3,995.56 | 160.06 | 120.90 | 2.34 | 15.8% | 18.6% | 3.0% |
| 10 | 1,146 | 1,653.13 | 2,067.93 | 211.14 | 232.17 | 1.25 | 3.1% | 2.7% | 33.2% |
| 11 | 308 | 2,124.63 | 3,863.22 | 217.39 | 215.46 | 1.73 | 4.5% | 2.6% | 20.5% |
| 12 | 428 | 1,430.85 | 2,044.16 | 150.12 | 153.32 | 1.45 | 5.8% | 6.1% | 32.5% |
| 13 | 1,089 | 1,707.93 | 3,452.68 | 192.10 | 181.34 | 1.28 | 6.2% | 5.0% | 14.5% |
| 14 | 1,036 | 2,548.94 | 4,832.29 | 78.23 | 67.21 | 1.45 | 6.6% | 2.9% | 12.5% |
| 15 | 1,057 | 2,137.00 | 2,863.16 | 194.92 | 208.97 | 1.16 | 6.1% | 6.8% | 17.8% |
| 16 | 3,132 | 6,305.94 | 8,510.53 | 240.87 | 240.26 | 1.30 | 7.1% | 4.9% | 6.9% |
| 17 | 1,009 | 2,550.60 | 5,648.97 | 161.52 | 175.76 | 1.16 | 4.7% | 6.8% | 7.1% |
| All | 9,920 | 3,341.85 | 6,070.61 | 194.96 | 194.46 | 1.42 | 6.2% | 5.4% | 13.9% |

# 5. Results and Discussion

Three types of impacts are tested in regards of new virtual LAND sales at the Sandbox. The first two tests focus on the new LAND sales effect and try to investigate the network externalities and/or increased supply effect of new virtual lands. The last one focuses on the external event, namely, the announcement of Meta (former Facebook) name change, and explores the impact from such event.

### 5.1 LAND Externalities

To motivate the DiD method, we first begin by computing the residuals from the regression of log prices on control variables and weekly fixed effects without Neari and Postt, which is the hedonic price regression without the DID variables. Next, the daily times series averages for the control and treatment groups are plotted in Figure 5. For the method to be appropriate, there should be a parallel trend between the two series pre-event and divergence post-event. The figure suggests that, on average, existing LANDs near the new parcels are sold at higher dollar prices than those further away. Combined with using announcements and exogenous events, the DiD method is justified.

**Figure 5: Average Residual Log Prices Around Announcement Dates**

This figure illustrates LAND transactions' average residual log prices around the announcement dates. The residuals are computed from the regression of log prices on control variables and weekly fixed effects without $Near_i$ and $Post_t$, which is the hedonic price regression without the DID variables. The daily times series averages for the control and treatment groups are plotted below, where the post-announcement periods are explicitly shaded.

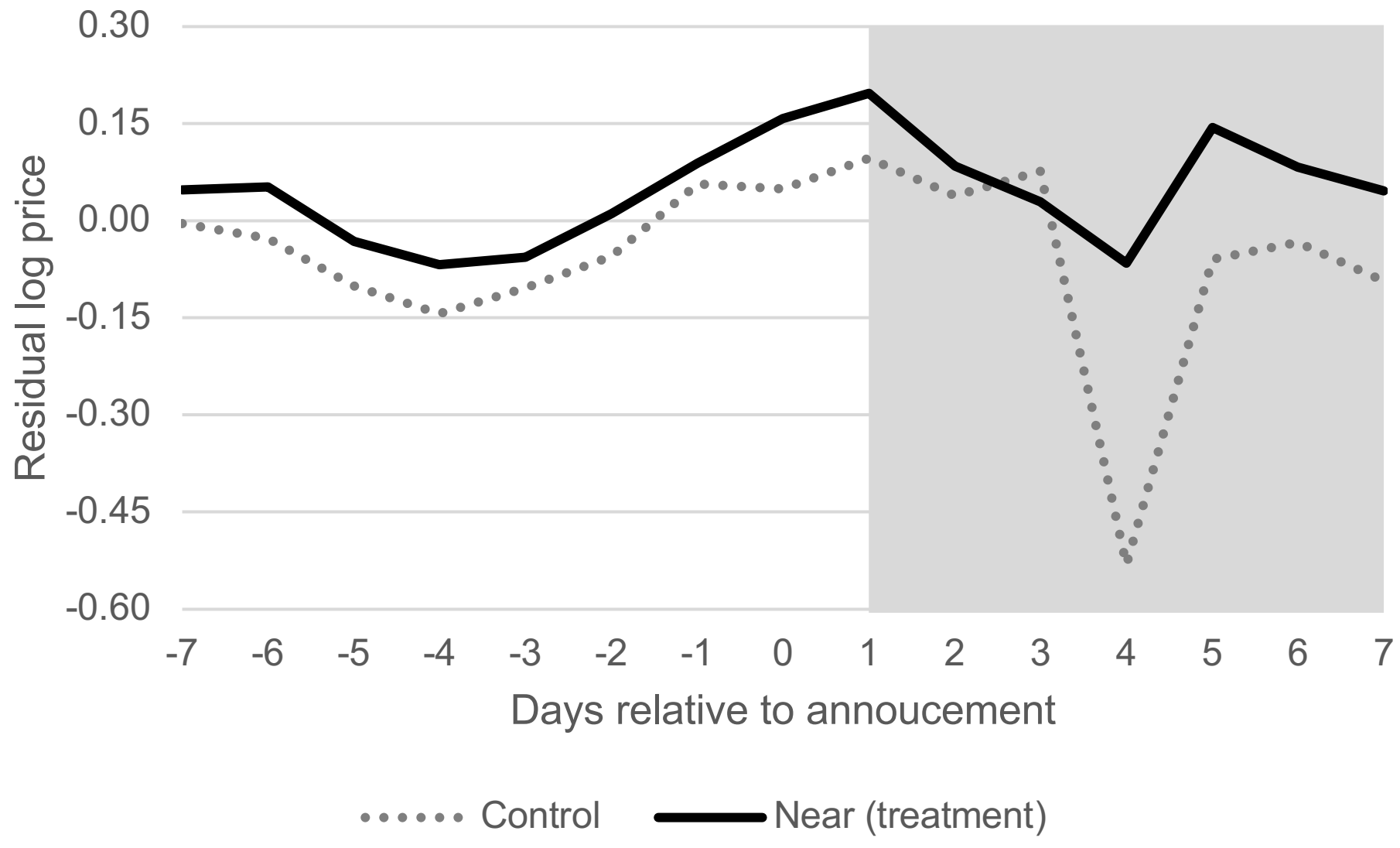

Table 3 reports our main regression results. Column 1 is the baseline model with weekly time fixed effects but no wave fixed effects. The log price specification allows the interpretation of $\beta_3$ as changes in percentage prices, and the estimated coefficient is positive as hypothesised. On average, the announcements of new public sales cause LAND prices near the new regions to increase by 8.4% in the seven days following the announcement. The increase is statistically significant at the 1% level.

Furthermore, it is robust to the addition of wave fixed effects that control for time-invariant heterogeneities of parcels (Column 2) and the utilisation of daily fixed effects rather than weekly (Column 3). Note that the log Bitcoin price is dropped because the data frequency is daily, so including daily fixed effects is more restrictive than the daily price. In the full-specification model (Column 4), where the choice of settlement cryptocurrency is included, the adjusted R-squared is increased, and the magnitude of $\beta_3$ is reduced to 7.7%. However, the coefficients remain statistically significant at the 1% level throughout.

The coefficients on SAND and wETH show that transactions settled in SAND are conducted at higher dollar price, while those settled in wETH at lower prices, compared to ETH which is the omitted category. The signs and magnitudes of the coefficients are consistent with Nakavachara and Saengchote (2022). The authors interpret this finding as evidence that users who paid with SAND which had appreciated sharply during the sample period would feel wealthier and thus are able to afford higher dollar prices.

**Table 3: Difference-in-Differences Regression Results (Discrete Treatment)**

This table reports the results of the difference-in-differences regressions of the log of LAND prices in dollars with discrete treatment. The indicator variable Neari is defined by the median Euclidean distance between new LANDs and the transacted LAND in each group of announcements. $Post_t$ is the indicator variable for transactions post-announcements. The standard control variables are the log of lot size, the indicator variable for premium LAND [omitted category is regular LAND], and the log of daily BTC price. Column 1 is the baseline model with weekly time fixed effects but no wave fixed effects. Column 2 includes both time and wave fixed effects. Column 3 replaces the weekly time fixed effects with the daily time fixed effects, dropping the daily BTC price variable. Column 4 also includes indicator variables for different tokens used to settle the sale [omitted category is ETH]. Stars correspond to the statistical significance level, with *, ** and *** representing 10%, 5% and 1%, respectively.

| | (1) | (2) | (3) | (4) |
|---|---|---|---|---|
| Post announcement | 0.111*** | 0.106*** | 3.721*** | 3.684*** |
| | (0.02) | (0.02) | (0.05) | (0.06) |
| Near new LAND (median) | 0.054*** | 0.030** | 0.033*** | 0.027** |
| | (0.01) | (0.01) | (0.01) | (0.01) |
| Post * Near | 0.084*** | 0.092*** | 0.084*** | 0.077*** |
| | (0.02) | (0.02) | (0.02) | (0.02) |
| Log(lot size) | 1.088*** | 1.079*** | 1.083*** | 1.071*** |
| | (0.02) | (0.02) | (0.02) | (0.02) |
| Premium LAND | 0.331*** | 0.455*** | 0.419*** | 0.420*** |
| | (0.03) | (0.03) | (0.03) | (0.03) |
| Log(BTC price) | 1.713*** | 1.650*** | | |
| | (0.14) | (0.14) | | |
| Paid in SAND | | | | 0.121*** |
| | | | | (0.03) |
| Paid in wETH | | | | -0.380*** |
| | | | | (0.02) |
| Time fixed effects | Week | Week | Day | Day |
| Wave fixed effects | No | Yes | Yes | Yes |
| Observations | 9,920 | 9,920 | 9,920 | 9,920 |
| Adj R-squared | 0.661 | 0.668 | 0.713 | 0.727 |

Table 4 reports our regression results for the continuous treatment variable. We replace the $Near$ variable with the log of distance to the new LANDs. We begin the first model by including the daily time and wave fixed effects, equivalent to Column 3 of Table 3. For continuous treatment, a lower distance means stronger treatment, so $\beta_3$ should be negative. As before, the full specification (Column 2) includes SAND and wETH variables. $\beta_3$ is negative and statistically significant for both specifications at the 1% level. For LANDs that are 1% closer (about two units, from the sample average in Table 1), their post-announcement prices are 3.2% higher. The coefficients' signs, magnitudes, and statistical significance for other variables remain similar. The discrete and continuous treatment results suggest positive externalities from increasing land supply, and its magnitude exceeds the negative price impact from increased supply.

### Table 4: Difference-in-Differences Regression Results (Continuous Treatment)

This table reports the results of the difference-in-differences regressions of the log of LAND prices in dollars with continuous treatment, defined as the log of distance between new LANDs and the transacted LAND. $Post_t$ is the indicator variable for transactions post-announcements. The standard control variables are the log of lot size and the indicator variable for premium LAND [omitted category is regular LAND]. Column 1 includes the daily time and wave fixed effects. Column 2 also includes indicator variables for different tokens used to settle the sale [omitted category is ETH]. Stars correspond to the statistical significance level, with *, ** and *** representing 10%, 5% and 1%, respectively.

| | (1) | (2) |
|---|---|---|
| Post announcement | 3.953*** | 3.903*** |
| | (0.08) | (0.09) |
| Log(distance) | -0.009 | -0.001 |
| | (0.01) | (0.01) |
| Post * log(distance) | -0.034*** | -0.032*** |
| | (0.01) | (0.01) |
| Log(lot size) | 1.082*** | 1.071*** |
| | (0.02) | (0.02) |
| Premium LAND | 0.418*** | 0.419*** |
| | (0.03) | (0.03) |
| Paid in SAND | | 0.122*** |
| | | (0.03) |
| Paid in wETH | | -0.382*** |
| | | (0.02) |
| Time fixed effects | Day | Day |
| Wave fixed effects | Yes | Yes |
| Observations | 9,920 | 9,920 |
| Adj R-squared | 0.713 | 0.726 |

### 5.2 The Supply Effect

From the discussion in Section 3.4, two competing forces could affect land prices post-announcements: the negative price impact from increased supply and the network externalities. We have established in Section 5.1 that the positive externalities exceed the negative price impact. In this section, we further delineate the two effects using the triple-differences method, which interacts with the DiD regression with another variable. Recall in Table 1 that some waves are announced jointly, while others independently. Waves that are jointly announced have more LAND available for sale. Using this variation in new LAND supply, we define a dummy variable $Multi_i$ to indicate announcements containing multiple waves, specifically the initial Summer Jam Waves (1-5) announced on 19 June 2021 and the Walking Dead Wave (1-4) announced on 21 August 2021.

$$\begin{aligned} y_{it} = \beta_1 Post_t + \beta_2 Treat_i + \beta_3 Multi_i \\ +\beta_4 Post * Multi_i + \beta_5 \; Post_t * Treat_i \\ +\beta_6 Post_i * Treat_i * Multi_i + \gamma' X_{it} + \alpha_w + \tau_t + \varepsilon_{it} \end{aligned} \quad (3)$$

The triple-differences model is outlined in Equation 3, where $Treat_i$ is used a generic name to indicate the treatment variable. The main coefficient of interest are $\beta_4$ and $\beta_6$, where $\beta_4$ represents a general decrease in LAND prices across The Sandbox, while $\beta_6$ represents a localised decrease near new parcels. We use the full specifications for both treatments and report the results in Table 5.

### Table 5: Triple-Differences Regression Results

This table reports the results of the triple-differences regressions of the log of LAND prices in dollars on the difference-in-differences variables and the additional interaction with $Multi_i$, the indicator variable for transactions that occur around announcements with multiple waves. For discrete treatment, the treatment variable is defined by the median Euclidean distance between new LANDs and the transacted LAND in each group of announcements. For continuous treatment, it is defined by the log of distance. $Post_t$ is the indicator variable for transactions post-announcements. The standard control variables are the log of lot size, the indicator variable for premium LAND [omitted category is regular LAND], and the indicator variables for different tokens used to settle the sale [omitted category is ETH]. Column 1 provides the full-specification result with discrete treatment and Column 2 with continuous treatment. Stars correspond to the statistical significance level, with *, ** and *** representing 10%, 5% and 1%, respectively.

| | (1)<br>Discrete | (2)<br>Continuous |
|---|---|---|
| Post announcement | 3.667*** | 3.928*** |
| | (0.07) | (0.10) |
| Treatment | 0.029*** | -0.001 |
| | (0.01) | (0.01) |
| Multi-wave | 2.217*** | 2.227*** |
| | (0.07) | (0.08) |
| Post * Multi-wave | -3.481*** | -3.681*** |
| | (0.09) | (0.12) |
| Post * Treatment | 0.106*** | -0.036** |
| | (0.02) | (0.02) |
| Post * Treatment * Multi-wave | -0.173*** | 0.014 |
| | (0.06) | (0.02) |
| Log(lot size) | 1.070*** | 1.071*** |
| | (0.02) | (0.02) |
| Premium LAND | 0.419*** | 0.419*** |
| | (0.03) | (0.03) |
| Paid in SAND | 0.121*** | 0.122*** |
| | (0.03) | (0.03) |
| Paid in wETH | -0.382*** | -0.382*** |
| | (0.02) | (0.02) |
| | | |
| Time fixed effects | Day | Day |
| Wave fixed effects | Yes | Yes |
| Observations | 9,920 | 9,920 |
| Adj R-squared | 0.727 | 0.726 |

Across both types of treatment, $\beta_4$ is negative and statistically significant at the 1% level, indicating wave size can exert a negative price pressure on existing LANDs. To the extent that all parcels have the same usage permissions (since there are no zoning restrictions in The Sandbox), there is less heterogeneity than in traditional real estate. However, proximity still matters, and if more parcels could be available for sale nearby, increasing local competition could decrease prices. For the discrete treatment, $\beta_6$ is -17.3%, statistically significant at the 1% level, and the DiD coefficient increases from 7.7% to 10.6%, consistent with the competing forces between the two effects discussed in Section 3.4. For the continuous treatment, β6 is positive as expected but statistically insignificant, and the DiD coefficient changes slightly from 3.2% to 3.6% per 1% change in distance. This difference in finding may result from non-monotonicity in the relationship between treatment and price.

Our results highlight the trade-off that developers face. On the one hand, more supply can diminish price, but on the other hand, network externalities can add value. The Sandbox natural experiment's unique research setting allows us to demonstrate these two forces' influence more explicitly.

### 5.3 The Meta Effect

The implicit assumption behind Rosen's (1974) hedonic pricing model is that users are rational, thus researchers can uncover the value of objectively measured characteristics from preferences revealed via transaction prices. This assumption is also implicit in the analysis by Goldberg, Kugler and Schär (2021) that Decentraland's users rationally assessed locations when auctioning for LANDs. However, the sample period for their analysis is 2018, when there were no sharp changes in cryptocurrency prices, including MANA, which is the cryptocurrency price of Decentraland, similar to The Sandbox's SAND.

In our sample period, Facebook announced a name change to Meta on 28 October 2021, leading to a sharp rise in metaverse-related cryptocurrencies such as MANA and SAND and their LAND NFTs. To illustrate this "Meta Effect", we plot the hedonic price index constructed from estimating Equation 1 with weekly fixed effects without DiD variables between January and December 2021 in Figure 6. The index starts from the first week of the year and reaches 11.1 in week 43 when Facebook made the announcement (depicted by the vertical line). By week 48, the index reached 83.0, almost 7.5 times higher than one month before. With such a rapid price rise, users might not make rational decisions and instead trade virtual real estate for capital gains.

**Figure 6: LAND Hedonic Price Index**

This figure illustrates the LAND hedonic price index constructed from estimating the hedonic price regression (Equation 1) with weekly fixed effects without DID variables between January and December 2021. The index starts from 1 at the beginning of the year's first week and reaches 11.1 in week 43, when Facebook announced it on 28 October 2021 (depicted by the vertical line). By week 48, the index reached 83.0, almost 7.5 times higher than one month before.

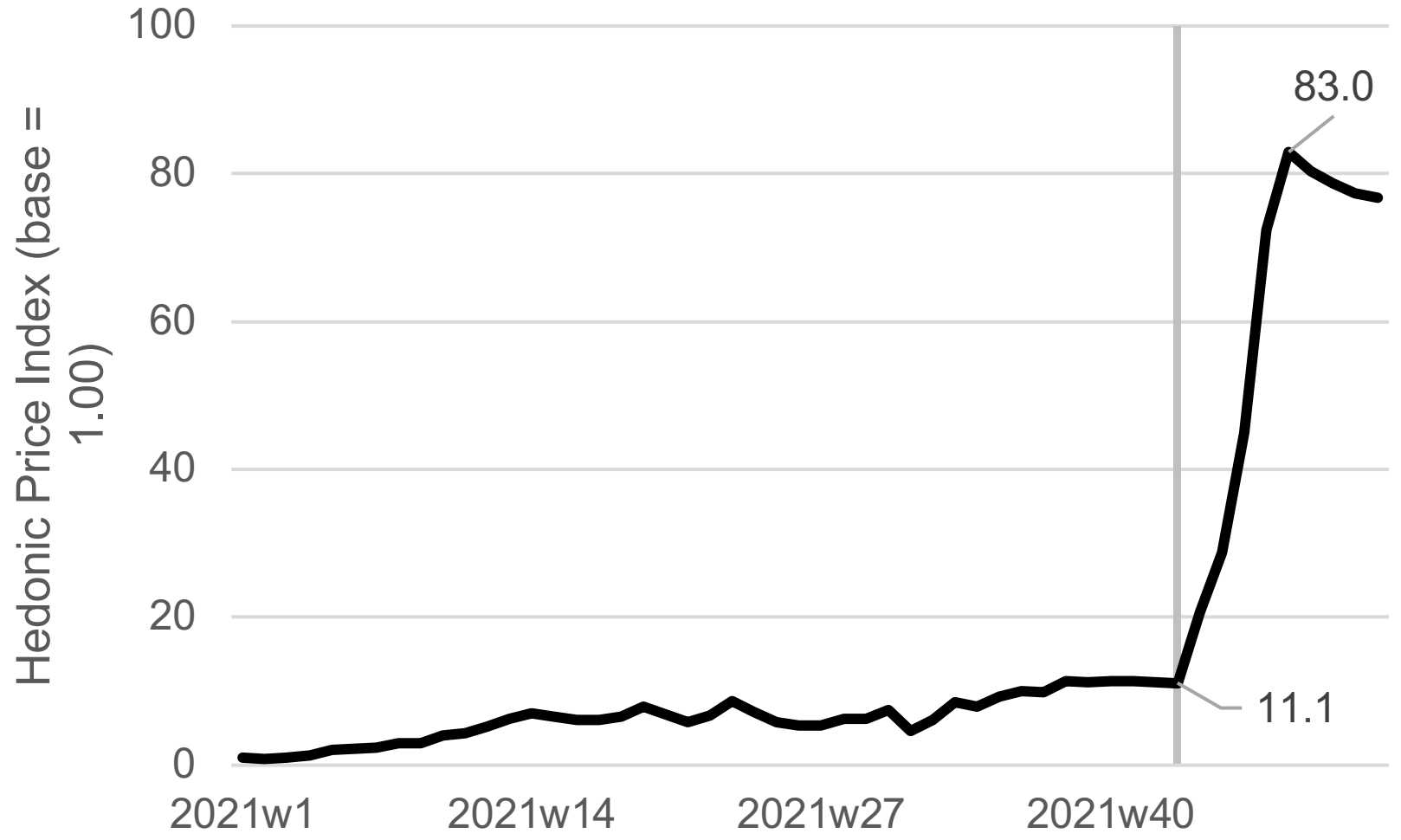

To investigate the Meta Effect, we partition the sample into two periods: before and after Facebook's announcement. The increased activity after the Meta announcement means more observations in the second period than in the first. We repeat the full specification DiD regressions using discrete and continuous treatments and report the results in Table 6. The results suggest that the network externalities are attributed to transactions before Facebook's announcement.

For the discrete treatment result (Column 1), the effect is an 18.6% increase in average price, while for the continuous treatment result (Column 3), the effect is a 4.8% increase for every 1% decrease in distance. The results are statistically significant at the 1% level, and the increase in magnitude is considerable. It is also worth noting that while the post-announcement partition is only ten weeks, there are more transactions in those weeks than in the preceding 42 weeks. The increased trading activity during this period corroborates the view of speculative trading, where users would be much less likely to capitalise externalities into their assessment. Thus, the network externalities are more substantial than previously estimated in Section 5.1.

### Table 6: The Meta Effect Results

This table reports the results of the difference-in-differences regressions of the log of LAND prices in dollars for announcements made before Facebook's name change on 28 October 2022 (Pre-Meta) and after (Post-Meta). For discrete treatment, the treatment variable is defined by the median Euclidean distance between new LANDs and the transacted LAND in each group of announcements. For continuous treatment, it is defined by the log distance. $Post_t$ is the indicator variable for transactions post-announcements. The standard control variables are the log of lot size, the indicator variable for premium LAND [omitted category is regular LAND], and the indicator variables for different tokens used to settle the sale [omitted category is ETH]. Columns 1 and 2 provide the full-specification results with discrete treatment Pre-Meta and Post-Meta. Columns 3 and 4 provide the full-specification results with continuous treatment Pre-Meta and Post-Meta. Stars correspond to the statistical significance level, with *, ** and *** representing 10%, 5% and 1%, respectively.

| | (1) | (2) | (3) | (4) |
|---|---|---|---|---|
| | Discrete Pre-Meta | Discrete Post-Meta | Continuous Pre-Meta | Continuous Post-Meta |
| Post announcement | 2.230*** | 1.147*** | 2.536*** | 1.228*** |
| | (0.08) | (0.03) | (0.11) | (0.09) |
| Treatment | -0.062*** | 0.046*** | 0.028*** | -0.011 |
| | (0.02) | (0.01) | (0.01) | (0.01) |
| Post * Treatment | 0.186*** | -0.007 | -0.048*** | -0.015 |
| | (0.04) | (0.02) | (0.02) | (0.02) |
| Log(lot size) | 1.018*** | 1.018*** | 1.073*** | 1.069*** |
| | (0.03) | (0.03) | (0.03) | (0.01) |
| Premium LAND | 0.333*** | 0.336*** | 0.383*** | 0.455*** |
| | (0.03) | (0.03) | (0.07) | (0.03) |
| Paid in SAND | 0.051** | 0.053** | 0.130*** | 0.107*** |
| | (0.02) | (0.02) | (0.04) | (0.03) |
| Paid in wETH | -0.310*** | -0.307*** | -0.332*** | -0.453*** |
| | (0.02) | (0.02) | (0.03) | (0.02) |
| Time fixed effects | Day | Day | Day | Day |
| Wave fixed effects | Yes | Yes | Yes | Yes |
| Observations | 4,722 | 5,198 | 4,722 | 5,198 |
| Adj R-squared | 0.536 | 0.825 | 0.534 | 0.825 |

## 6. Conclusion

Network externalities have crucial impacts to value in connected systems. This idea also applies to land: when new land becomes available, the network of existing land improves due to better connection, so better connected land whould have a higher value. Network externalities are intuitive and appealing to developers and policymakers, but documenting their importance is empirically challenging. In traditional real estate, changes in land supply are often accompanied by other developments that could confound the estimated value of network externalities.

In this study, we use The Sandbox metaverse's virtual real estate market to isolate the value of network externalities. By exploiting unanticipated announcements of new LAND sales as natural experiments that exogenously change land supply and the difference-in-differences method, we find that LAND prices in neighbouring areas can increase as much as 8.4% in the seven days following the announcement, consistent with the network externalities effect. Furthermore, land sale waves have different sizes, so we use the triple-differences regression to demonstrate that new supply can have a competing, adverse effect on existing LANDs. Finally, we show that the speculative frenzy can weaken the incentive to make a rational decision as the value of network externalities is not capitalised into transaction prices post-Meta announcement.

While Rosen's (1974) hedonic pricing model provides a framework to attribute sources of value, limitations of traditional real estate make causal inferences challenging. While our analyses are conducted on virtual real estate, the results are consistent with the theories that apply to physical real estate. The results from The Sandbox's natural experiments in this study will shed light to the evidence-based policymaking and illustrate the benefits of using virtual worlds to conduct real-world policy experiments.